# Optical amplification of surface plasmon polaritons in a graphene single layer integrated with a random grating

Abbas Ghasempour Ardakani[1,*], Peymaneh Rafieipour[1]

[1]*Department of Physics, College of Science, Shiraz University, Shiraz 71454, Iran*
*Corresponding author: aghasempour@shirazu.ac.ir*

**Abstract**

In this paper, we design and simulate a terahertz (THz) controllable active plasmonic waveguide structure based on a single graphene layer that is placed on a random silicon grating substrate. Optical gain in the proposed THz active plasmonic waveguide structure is provided by the stimulated emission process in the photoexcited graphene monolayer that leads to the amplification of surface plasmon polariton (SPP) waves. We use a random grating substrate to introduce Anderson localization of the SPP waves propagating through the graphene monolayer to enhance their optical amplification at resonant frequencies. It is shown that the enhancement factor of the resonant peaks corresponding to the graphene SPPs can be as high as 175. We also analyze their corresponding field intensity distributions along the graphene monolayer and find out that their intensities and localization positions are different from each other. By investigating the pump dependent and temperature dependent variations of the transmittance of the structure, it is shown that the resonant peak frequencies are blue-shifted by increasing the temperature and the external pump intensity. Also, we show that increasing the ambient temperature by 60 K can dramatically reduce the output amplified intensity by a factor of 70. This property of the proposed graphene-based THz plasmonic waveguide structure makes it useful in temperature sensing applications and on/off switchable laser devices.

**1.Introduction**

Graphene is a single two-dimensional layer of $sp^2$-hybridized carbon atoms arranged in a hexagonal lattice. Due to its high electron mobility, gapless and linear dispersion near the Dirac points and tunable electrical and optical properties, using graphene is extremely desirable in many applications including photodetectors, transistors, modulators, biosensors, smart windows and solar cells [1]. Studies on graphene surface plasmon polaritons and their associated device technologies are rapidly evolving, because of their higher electromagnetic confinement, lower ohmic loss and longer propagation lengths in comparison with those of metal surface plasmon polaritons [2]. Surface plasmons are collective electron oscillations at metal/dielectric or graphene/dielectric interfaces and their coupling with electromagnetic photons results in the formation of surface plasmon polariton (SPP) waves that provide a high electromagnetic field confinement and optical intensity at sub-wavelength scales [3]. In a recent theoretical paper, Anderson localization of surface plasmons was demonstrated in a single graphene layer deposited on a random silicon grating substrate [4]. M. Sani et. al. showed that the excited graphene surface plasmons become localized in some random positions through the graphene single layer, when the silicone trench width varies randomly. Also, the field intensity distributions corresponding to the localized graphene surface plasmons revealed that they are spatially localized in certain regions along the graphene layer which act like a resonator and their filed intensity strongly enhances at the localization sites. Very recently, temperature tunable Anderson localization of propagating graphene SPP waves was demonstrated in a disordered plasmonic structure composed of a graphene monolayer placed on a random grating substrate composed of InAs [5]. A. G. Ardakani et. al. used the temperature dependent electric permitivitty of InAs to tune the transmission bands of the random plasmonic structure. Their results showed that the resonance peaks corresponding to the graphene SPPs are blue-shifted, by increasing the ambient temperature from 700 K to 850 K. By calculating the localization lengths of the disordered plasmonic structure, they showed that the number of localized plasmonic states increases with increasing the disordered level.

The population inversion and THz stimulated emission in a graphene single layer was studied in many theoretical researches. It was shown that the photogeneration of electron-hole pairs in graphene under sufficiently strong optical pumping and their radiative recombination results in the stimulated photon emission in the THz range of frequencies [6]. This phenomenon was the physical basis employed in designing many graphene-based lasers operating in the THz range of frequencies [7]. Also, the amplification of surface plasmons propagating along optically pumped single-graphene layer and multiple-graphene layer structures placed on an unpatterned substarte was demonstrated [8]. A. A. Dubinov et. al. investigated the effects of temperature, quasi-Fermi energy, substrate refraction indices and the number of graphene layers on the surface plasmon absorption coefficient (plasmon gain) and demonstrated that the plasmon gain in a single-graphene layer structure is markedly higher than the gain of the electromagnetic modes in a dielectric counterpart. Very recently, I. M. Moiseenko et. al. proposed a dual grating-gate graphene-based structure with an asymmetric unit cell to excite radiative and nonradiative plasmon modes by normally incident THz electromagnetic wave and study their amplification by graphene [9]. With the aim of optimizing the excitation and amplification of surface plasmons, they used an integral equation method to solve the full system of the Maxwell equations and investigate the effects of the structural parameters as well as the momentum relaxation time on the plasmon absorbance coefficient and the transformation coefficient of normally incident electromagnetic wave into propagating plasmons. Experimentally, THz time-domain spectroscopy measurements demonstrated the stimulated THz photon emission with excitation of SPPs in photoexcited monolayer graphene [10]. T. Watanable et. al. studied an intrinsic monolayer graphene on a SiO2/Si substrate under optical pump with a femtosecond-infrared laser pulse, THz probe pulse and optical probe (femtosecond-pulsed fiber laser) measurements. They detected an intense THz probe pulse as a result of the excitation of SPPs and showed that the obtained gain enhancement factor can be more than 50. Regarding the optical amplification by a graphene monolayer, it is a promising material for realizing THz photonic, electronic and optoelectronic devices, including graphene based amplifiers, lasers and random lasers.

In this paper, we combine the gain of graphene with the Anderson localization of graphene SPPs to significantly enhance the intensity of SPP waves at resonant frequencies of the random plasmonic waveguide structure. Since structural artifacts are unavoidable in fabricating the photonic industrial materials and they induce multiple scattering events due to the randomness, studying disordered structures is more desirable from a practical point of view. In addition, due to Anderson localization effect, local cavities can be formed in the random structures which induce more amplification. Hence, we suggest a disordered grating structure made of silicon as the substrate of a garphene monolayer and introduce the disorder by changing the silicone trench width, randomly. By calculating the transmittance for the SPP waves propagating along the graphene monolayer, it is shown that the transmittance becomes much larger than one at resonant frequencies due to the presence of optical gain and localized random cavities. Resonant frequencies in the transmission spectrum are blue-shifted by increasing the temperature and the external pump intensity. The obtained results suggest that the enhancement factor of the resonant peaks corresponding to the graphene SPPs can be as high as 175. On the contrary, increasing the ambient temperature from 300 K to 360 K can dramatically reduce the transmission intensity by a factor of 70. This property of the proposed graphene-based THz plasmonic waveguide structure makes it useful in temperature sensing applications and on/off switchable laser devices.

## 2.Structure design and simulation method

The surface electrical conductivity of graphene in the THz range of frequencies is described by the Kubo formula, according to the relation [8]:

$$\sigma = (\frac{e^2}{4\hbar})\left\{\frac{8k_BT\tau}{\pi\hbar(1-i\omega\tau)}\ln\left[1+\exp(\frac{\varepsilon_F}{k_BT})\right]+\tanh\left(\frac{\hbar\omega-2\varepsilon_F}{4k_BT}\right)-\frac{4\hbar\omega}{i\pi}\int_0^\infty\frac{G(\varepsilon,\varepsilon_F)-G(\hbar\omega/2,\varepsilon_F)}{(\hbar\omega)^2-4\varepsilon^2}d\varepsilon\right\} \quad (1)$$

where e is the electron charge, ℏ is the reduced Plank constant, $k_B$ is the Boltzmann constant, T is the temperature, τ is the electron and hole momentum relaxation time and $\varepsilon_F$ is the quasi-Fermi energy. The first term in the graphene conductivity describes the intraband electron-photon scattering and the last two terms correspond to interband transitions of charge carriers in graphene. Also, the function $G(\varepsilon,\varepsilon_F)$ is defined as follows [8]:

$$G(\varepsilon,\varepsilon') = \frac{\sinh(\varepsilon/k_BT)}{\cosh(\varepsilon/k_BT)+\cosh(\varepsilon'/k_BT)} \quad (2)$$

Under the pulse photoexcitation, $\varepsilon_F$ varies as a function of the pump intensity via the following relation [6]:

$$\varepsilon_F = 12\alpha\hbar^2v_F^2\frac{\tau_R I_\Omega}{\hbar\Omega k_BT} \quad (3)$$

where $\alpha=(e^2/\hbar)\sim 1/137$ is the fine structure constant, $v_F$ is the Fermi velocity, $\tau_R$ is the characteristic recombination time of the photogenerated electrons and holes, $I_\Omega$ and $\hbar\Omega$ are the intensity and photon energy of the incident optical radiation, respectively. The optical properties of graphene are described by its electrical permittivity as [5]:

$$\varepsilon_g = 2.5 + i\frac{\sigma}{\omega\varepsilon_0 d_g} \quad (4)$$

where $\varepsilon_0$ is the vacuum permittivity and $d_g$ is the thickness of the single graphene layer. Fig. 1 displays the frequency dependence of the real part of the graphene permitivity for different values of the pump intensity. The simulation parameters are set as T=300 K, τ=0.67 ps, $v_F=10^6$ m/s, $\tau_R=10^{-7}$s, ℏΩ=0.8 eV and $d_g$=1 nm.

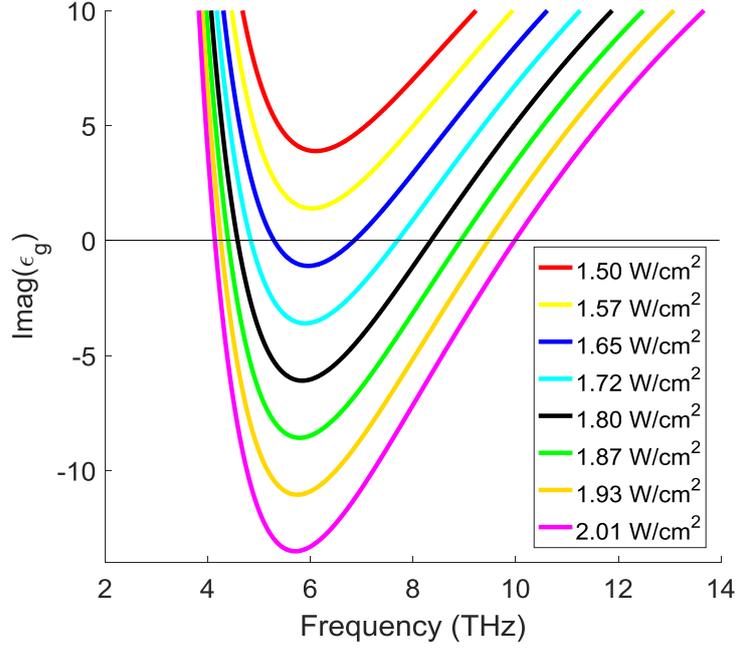

**Figure 1. Imaginary part of the graphene electrical permittivity as a function of frequency for different values of the pump intensity.**

We observe that imaginary part of the graphene permitivity becomes negative, when the pump intensity exceeds a certain value. Hence, the single graphene layer can act as a gain material and amplify an electromagnetic wave propagating along the grpahene sheet with the frequencies located between 4 THz to 10 THz. In addition, the absolute value of the negative real part of the graphene conductivity and the negative imaginary part of the graphene electrical permittivity increases, when the pump intensity increases. This is equivalent to the increase of gain in graphene as a result of the increase in the pump intensity. Also, the width of frequency range in which the real part of the graphene conductivity and the imaginary part of the graphene electrical permittivity become negative increases by increasing the pump intensity. It implies that optical amplification is possible in larger frequency range with the increase of pump intensity exciting the graphene layer.

As it is shown schematically in Fig. 2, the graphene-based THz plasmonic waveguide structure proposed in this paper consists of a single graphene layer deposited on a random silicon grating substrate surrounded by air. We assume that the structure is infinite in the z direction and SPP waves propagate along the graphene monolayer in the x direction.

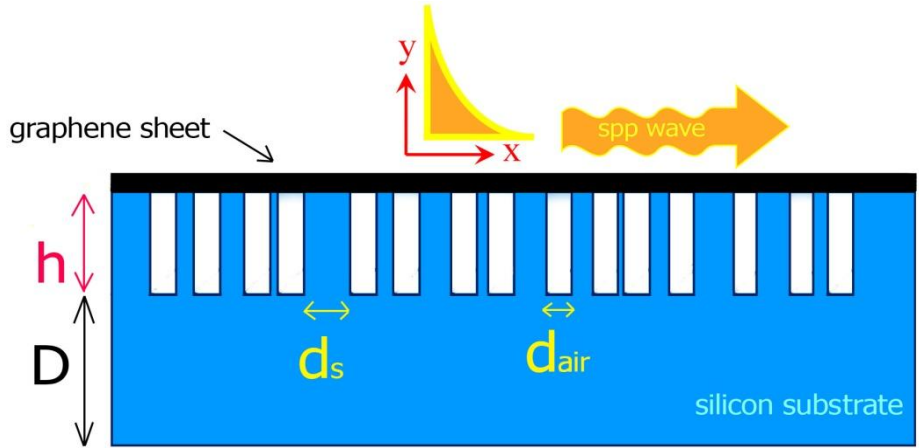

**Figure 2.** A schematic view of the proposed graphene-based plasmonic waveguide structure. The graphene monolayer is placed on a random silicon grating substrate and SPP waves with frequencies in the range of 4 THz to 10 THz are supposed to inject through the graphene sheet from the left. The width of the air trenches, width of the substrate grooves, height of the substrate grooves and thickness of the substrate are denoted by $d_{air}$, $d_s$, h and D, respectively.

We use two-dimensional simulation by the finite element method software (COMSOL Multiphysics) to investigate the proposed structure, numerically. In the simulation, we use a numeric port to excite SPPs in the graphene layer [11]. Furthermore, the grating structure is partially random in the sense that the grating line width varies randomly, but the width of the air grooves is constant. The depth and width of the air grooves, thickness of the substrate and width of the grating lines are denoted by h, $d_{air}$, D and $d_s$, respectively. We assume that the width of the grating lines is obtained from the relation $d=d_{avg}(1+w)$ where $d_{avg}$ is a constant and w is a random number distributed uniformly in the range [-q,q]. In other words, $d_{avg}$ is the average width of the grating lines while q corresponds to the disorder strength. In our simulation, we take $d_{avg}$=50 nm and q=0.9. Other simulation parameters including the width of the air trenches, thickness of the substrate, depth of the air grooves and the thickness of the graphene monolayer are 50 nm, 100 nm, 25 nm and 1 nm, respectively. In addition, the electrical permittivity of silicon is set to 11.8 in the simulations.

### 3. Results and discussion

Figure 3 illustrates the pump dependent variation of the transmission spectra, when the graphene monolayer is excited by a pump light with the wavelength of 1550 nm (photon energy of 0.8 eV) at room temperature.

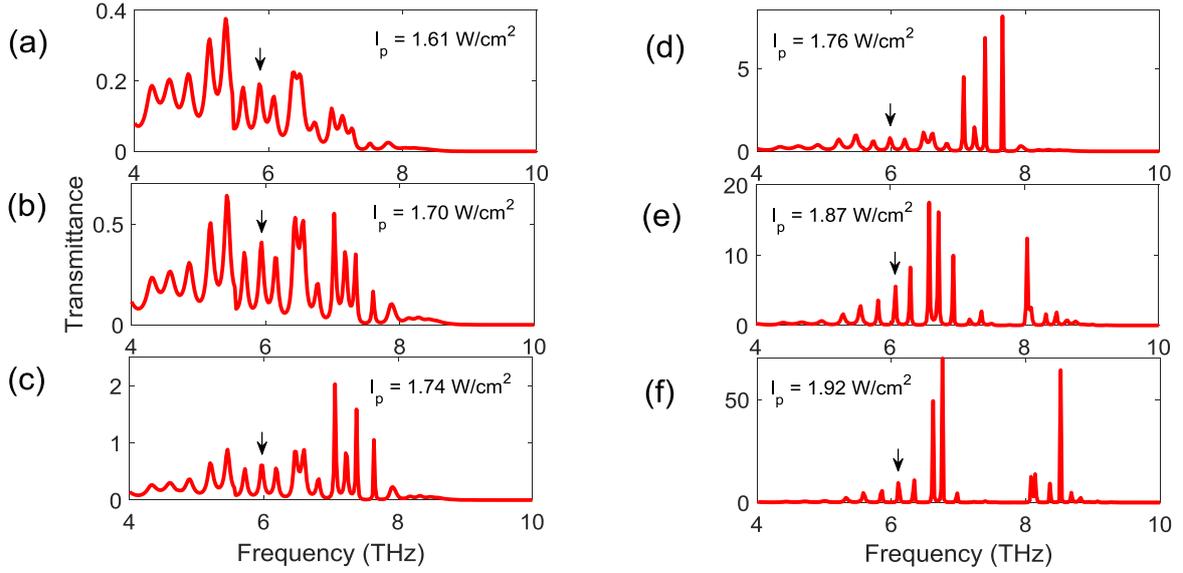

**Figure 3.** Pump dependent variation of the transmission spectrum: figures (a) to (f) exhibit the transmission spectra corresponding to the pump intensities of 1.61 W/cm$^2$, 1.70 W/cm$^2$, 1.74 W/cm$^2$, 1.76 W/cm$^2$, 1.87 W/cm$^2$ and 1.92 W/cm$^2$, respectively. An arrow is used for denoting the spectral position of a typical resonance peak.

There are several noticeable points in the results presented in Fig. 3. First, we observe that the transmission intensity increases by a factor of 175, when the pump intensity increases from 1.61 W/cm$^2$ to 1.92 W/cm$^2$. It is expected because the gain of graphene increases by increasing the pump intensity, as shown in Fig. 1 (a), and the propagating SPPs obtain more amplification before they leave the graphene sheet. It is worth mentioning that there are several random local cavities along the graphene sheet as a result of the Anderson localization of graphene SPPs. Hence, the advantageous of using Anderson localization of graphene SPPs is that it enhances the intensity of graphene SPPs at the local cavities which in turn results in the increase of the transmittance of the random structure. As the second remarkable point in Fig. 3, the linewidth of the resonant peaks are very broad and the transmission intensity is weak at low pump intensities. Also, we observe that there exist spectral overlaps between the adjacent peaks at low pump intensities. It is because the small amount of gain available at low pump intensities is insufficient for the resonant frequencies of the structure to overcome the loss and leave the structure, independently. The frequencies of these resonant peaks at low pump intensities are very close to the resonant frequencies of the passive random structure, as it is demonstrated in Fig. 4 for the pump intensity of 1.61 W/cm$^2$.

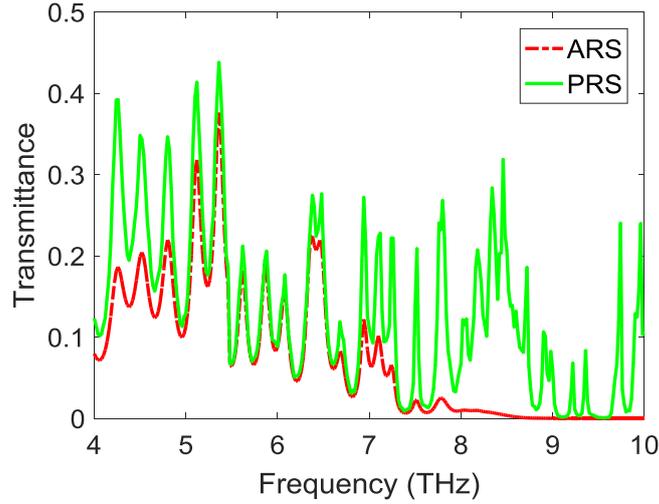

**Figure 4.** Comparison between the transmission spectra for the structure at the pump intensity of 1.57 W/cm$^2$ (ARS) and the structure in which the imaginary part of permittivity of graphene is ignored (PRS). Other simulation parameters are the same in two curves.

It should be noted that to obtain the resonant peaks in the structure, we ignore the imaginary part of the electrical permittivity of graphene. In this case where there is no loss or gain in the graphene layer, we named the structure as PRS that is the abbreviation of "passive random structure". One can observe that the frequencies of the resonant peaks in the active random structure (ARS), where the gain of graphene is included in the analysis, are nearly the same with the resonant frequencies of the passive random structure (PRS) in the range of 4 THz to 7.5 THz. The quality factors of the random local cavities ($v_{resonant}/\Delta v$) depend on the quasi-Fermi energy as well as the resonance frequency and vary from 40 to 190. In these random cavities the optical loss results from only the output coupling because we neglect the optical loss of the graphene layer in the calculation of the resonant peaks. It is worth mentioning that the disappearance of the resonant peaks with the frequencies higher than 7.5 THz in the presence of the gain of graphene monolayer at low pump intensity (red curve in Fig. 4) is due to high absorbance of graphene layer for frequencies located in this range as shown in Fig. 1. At high pump intensities when more optical gain are available for the resonant frequencies of the structure, the resonant peaks are much narrower, more intensive and are resolved better from each other (Fig. 3). In addition, the resonant peaks with higher frequencies can gain the required amplification and appear at larger values of the pump intensity. As a result, the number of amplified peaks in the transmission spectra increases by increasing the pump intensity. It is because the frequency range in which the graphene acts as the gain material increases by increasing the pump intensity (Fig. 1). The third point extracted from Fig. 3 is that the transmission spectrum is blue-shifted by 0.24 THz, when the pump intensity increases from 1.61 W/cm$^2$ to 1.92 W/cm$^2$. We mark a typical amplified peak with an arrow to better visualize the spectral shift of the transmission spectrum. The first reason for this pump-dependent behavior is that the gain spectrum of graphene exhibits a blue-shift of 0.36 THz to higher frequencies, when the pump intensity increases (Fig. 1). The second reason is that the resonance frequencies resulted from Anderson localization of graphene SPPs are blue-shifted with the increase of the quasi-Fermi energies as a result of increasing the pump intensity. One may think of it as a simple method for tuning the transmittance of the proposed graphene-based plasmonic waveguide structure in the THz range of frequencies.

Figure 5 (a) shows the spectrally integrated transmittance versus the pump intensity for the proposed graphene-based plasmonic waveguide structure. We clearly observe that there is a threshold pump intensity at which the spectrally integrated transmission intensity starts to increase very rapidly with the increase of the pump intensity.

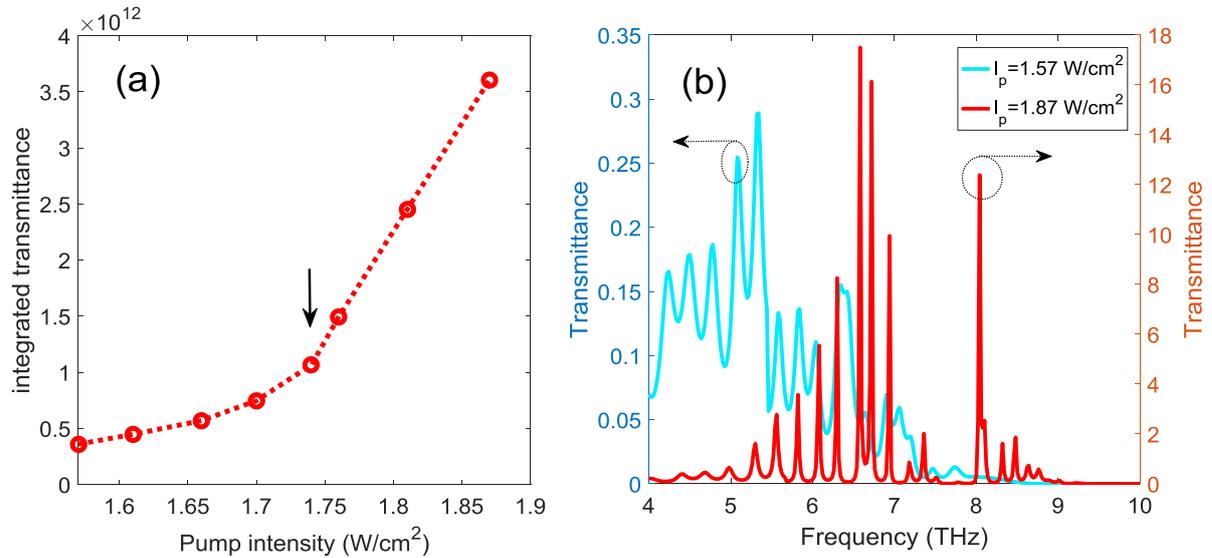

**Figure 5.** (a) Spectrally integrated transmittance versus the pump intensity, (b) The transmission spectra at two different pump intensities below and above the threshold (1.74 W/cm$^2$).

One can estimate the threshold pump intensity around 1.74 W/cm$^2$, from the obtained data presented in Fig. 5 (a). This threshold indicates the pump intensity above which more slope efficiency occurs and more efficient amplification is obtained.

Therefore, the graphene monolayer deposited on a random grating substrate acts as an amplifier in which the gain is provided by the graphene layer and the Anderson localization of graphene SPPs enhances the amplification process. Two transmission spectra below and above the threshold corresponding to the pump intensities of 1.57 W/cm$^2$ and 1.87 W/cm$^2$ are shown in Fig. 5 (b), respectively. By comparing their intensities and linewidths, the narrow amplified peaks with linewidths as low as 0.02 THz are observed in the transmittance of the graphene monolayer deposited on the random silicon grating, at high pump intensities well above the threshold.

Figures 6 (a-k) show the field intensity distributions corresponding to the amplified peaks seen in the transmission spectrum, when the pump intensity is 1.87 W/cm$^2$. From top to down, Fig. 6 (a) to Fig. 6 (k) correspond to the lasing peaks in Fig. 3(e) with frequencies of 5.56 THz, 5.82 THz, 6.08 THz, 6.3 THz, 6.58 THz, 6.72 THz, 6.94 THz, 7.36 THz, 8.04 THz, 8.32 THz and 8.48 THz, respectively. The inset displays a magnified image of the field intensity distribution corresponding to the resonant peak with the frequency of 8.04 THz.

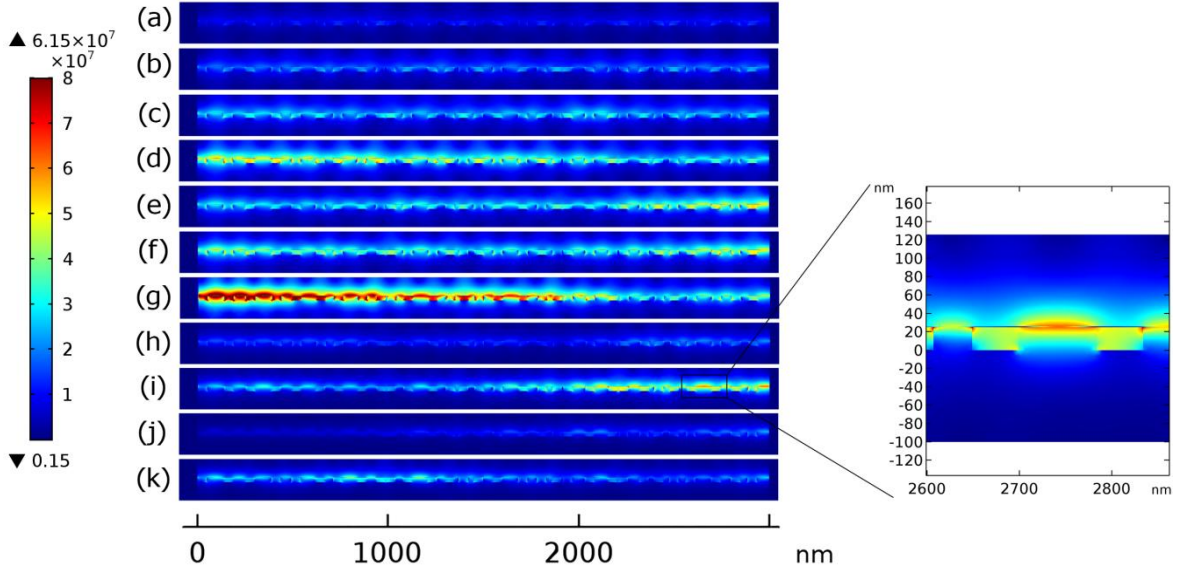

**Figure 6.** Field intensity distributions (in unit of V/m) of the resonance peaks appear in the transmission spectrum, when the pump intensity is 1.87 W/cm². Figures (a) to (k) correspond to the intensity distributions of the resonance peaks with the frequencies of 5.56 THz, 5.82 THz, 6.08 THz, 6.3 THz, 6.58 THz, 6.72 THz, 6.94 THz, 7.36 THz, 8.04 THz, 8.32 THz and 8.48 THz, respectively. The inset shows a magnified image of the field confinement and SPP localization to the graphene monolayer.

It is clearly seen in Fig. 6 that the propagating SPP waves are spatially localized through the graphene sheet in some positions along the random grating substrate. Also, their intensity and localization positions are different from each other. For example, field intensity distribution corresponding to the resonant peak with the frequency of 6.3 THz is confined to the left side of the structure [Fig. 6(d)], in contrast to the field intensity distribution corresponding to the resonant peak with the frequency of 6.58 THz that is confined to the right part of the structure [Fig. 6 (e)]. In addition, the resonant peak with the frequency of 6.94 THz has the strongest field confinement among the other resonant peaks [Fig. 6 (g)]. It seems important to mention here that the transmittance corresponding to the resonant peaks depicted in Fig. 5 (b) is the output intensity of SPP waves that leave the structure and is different from the field intensities of the hot-spots shown in Fig. 6 which are created as a result of the scattering of SPP waves and their localization along the random grating substrate.

Next, we change the ambient temperature and investigate its effects on the transmission spectra. Figure 7 illustrates the temperature dependent variation of the transmission spectra when the pump intensity is 1.92 W/cm².

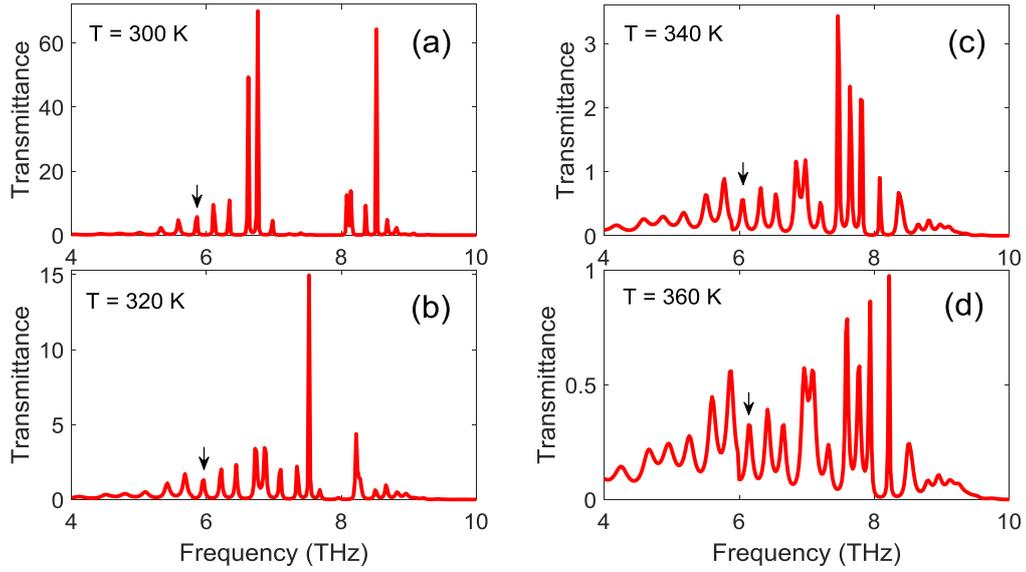

**Figure 7.** Temperature dependent variation of the transmission spectrum: Figures (a) to (d) exhibit the transmission spectra corresponding to different ambient temperatures of 300 K, 320 K, 340 K and 360 K, respectively. An arrow is used for denoting the spectral position of a typical resonance peak. The pump intensity is constant and equals 1.92 W/cm$^2$.

We observe in Fig. 7 that the transmittance reduces by a factor of 70, when the temperature increases by 60 K. Since the linewidth of the resonant peaks is increased and they overlap at high temperatures, we come to this conclusion that the transmittance significantly decreases at the high temperature of 360 K for the fixed intensity of 1.92 W/cm$^2$. It should be noted that at frequencies at which the transmittance is smaller than one there is no amplification. To determine why the intensity of the resonant peaks decreases with increasing the temperature, we plot the imaginary part of graphene permittivity at different temperatures 300, 320, 340 and 360 K and display the corresponding results in Fig. 8. One can see that when temperature increases from 300 to 360 K, the absolute value of the imaginary part of graphene permittivity decreases for the frequencies in the range 4 to 6.5 THz. Furthermore, the frequency range at which the imaginary part of graphene permittivity is negative shifts to higher frequencies. However, the absolute value of the imaginary part of graphene permittivity increases for the frequency in the range 6.5 to 9 THz. In addition, according to Eq. (3), when temperature increases at a fixed pump intensity, the quasi-Fermi energy decreases resulting the red-shift of the resonant wavelength (The details for the red-shift of the resonant peaks with decrease of quasi-Fermi energy are not displayed here). Consequently, the combination of these effects causes the transmittance of the amplified peaks to decrease with temperature increasing, as observed in Fig. 7 (a-d).

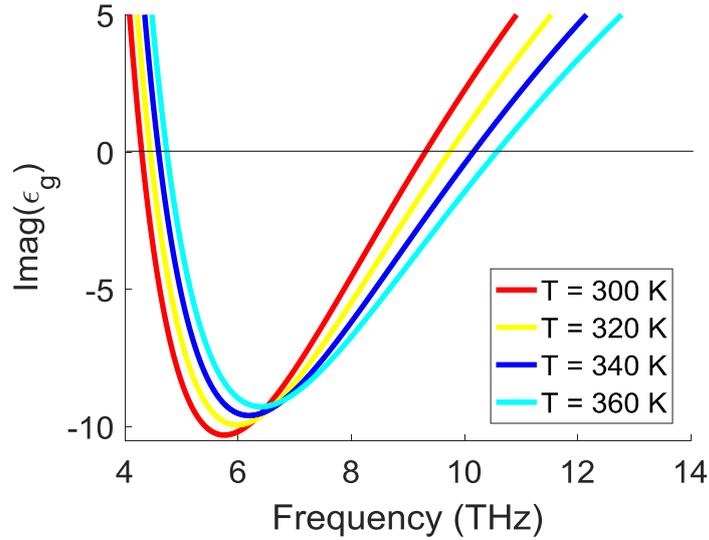

Figure 8. Terahertz frequency dependence of the imaginary part of the graphene electrical permittivity, corresponding to different ambient temperatures of 300 K, 320 K, 340 K and 360 K. The pump intensity is 1.92 W/cm$^2$.

Temperature-dependent characteristics of the proposed graphene-based THz plasmonic waveguide structure can be used in the fast on-off switchable THz devices and in remote temperature sensors. In addition, it is clearly observed that changing the temperature can be regarded as a facile route for tuning and controlling the transmittance and the resonant peaks. As it is evident from the small arrow located on top of a typical resonant peak, the transmittance is blue-shifted by approximately 0.3 THz, when the temperature of the graphene monolayer is increased by 60 K. This effect is due to the blue-shift of frequency range where the imaginary part of permittivity is negative by increasing the temperature, as shown in Fig. 8.

Our results in this paper confirm that a random grating can be used to enhance the amplification of SPPs in a graphene monolayer at resonant frequencies of the ransom structures like a Fabry-Perot cavity. It is well-known that there are many resonant peaks in random structures. Therefore, we can amplify the SPP waves in a large number of wavelengths. In addition, the resonant frequencies can be changed with variation of Fermi energy and the structure parameters such as air trenches, width of the substrate grooves and height of the substrate grooves. We expect that the experimental realization of the proposed structure in this paper can be found applications in the integrated optics. It should be emphasized that if the radiative recombination of electron-holes occurs in the optical pumped graphene and the resulted radiation can excite the SPPs in the graphene layer, we expect that the proposed structure in this paper acts as a plasmonic random laser or a spasers without any input signal.

**Conclusions**

In conclusion, we have designed and simulated a terahertz controllable plasmonic waveguide structure based on a graphene monolayer deposited on a random grating substrate made of silicon. The simulation results show that the enhancement factor of the resonant peaks corresponding to the graphene SPPs can be as high as 175. Furthermore, the transmittance undergoes a shift to higher frequencies as a result of the increase in the pump intensity. By analyzing the field intensity distributions corresponding to each resonant peak at a constant pump intensity, it is shown that the intensity and localization position corresponding to each resonant peak are different from the others. Finally, we have studied the temperature dependent variation of the transmission spectrum. It has been demonstrated that when the ambient temperature increases, amplification decreases and the frequencies of resonant peaks are blue-shifted. These properties of the proposed graphene-based THz plasmonic waveguide structure can be applied in temperature sensing applications and on/off switchable laser devices.